\documentclass[runningheads,a4paper]{llncs}

\usepackage{setspace}

\usepackage{amssymb}
\usepackage{graphicx}
\usepackage{amsmath}
\usepackage{color}
\usepackage{rotating}

\begin{document}
\mainmatter  
\title{An Integer Programming Formulation of the Minimum Common String Partition problem}
\titlerunning{An IP Formulation of the MCSP problem}
\author{S. M. Ferdous\inst{1 \and 2} \and M. Sohel Rahman\inst{1}}

\institute{A$\ell$EDA Group, Department of CSE, BUET, Dhaka-1000, Bangladesh\\
\and Ahsanullah University of Science and Technology (AUST), Dhaka-1208, Bangladesh}
\maketitle

\begin{abstract}
We consider the problem of finding a minimum common partition of two strings (MCSP). The problem has its application in genome comparison. MCSP problem is proved to be NP-hard. In this paper, we develop an Integer Programming (IP) formulation for the problem and implement it. The experimental results are compared with the previous state-of-the-art algorithms and are found to be promising.
\end{abstract}

\section{Introduction}

String comparison is one of the major problems in computer science with extensive applications in different areas that includes genome sequencing, text processing and compression. In this paper, we address the problem of finding a minimum common partition (MCSP) of two strings. MCSP is closely related to genome arrangement, an important field in Computational Biology.
More detailed study of the application of MCSP can be found at \cite{goldstein}, \cite{chen} and \cite{chrobak}.

In the MCSP problem, we are given two \emph{related} strings $(X,Y)$. Two strings are related if every letter appears the same number of times in each of them. Clearly, two strings have a common partition if and only if they are related. So, the length of the two strings are also the same (say, $n$). A partition of a string $X$ is a sequence $P = (B_1,B_2, \cdot\cdot\cdot ,B_c)$ of strings whose
concatenation is equal to $X$, that is $B_1B_2 \cdot\cdot\cdot B_c = X$. The strings $B_i$ are called
the blocks of $P$. Given a partition $P$ of a string $X$ and a partition $Q$ of a string
$Y$, we say that the pair $\pi = <P,Q>$ is a common partition of $X$ and $Y$ if $Q$ is a
permutation of $P$. The minimum common string partition problem is to find a
common partition of $X$, $Y$ with the minimum number of blocks, that is to minimize $c$. For example, if $(X,Y)$ = (``ababcab'',``abcabab''), then one of the minimum common partition sets is $\pi = $\{``ab'',``abc'',``ab''\} and the minimum common partition size is 3. The restricted
version of MCSP where each letter occurs at most $k$ times in each input string, is denoted by $k$-MCSP.

In this paper, we present an Integer Programming formulation for the MCSP problem. In particular, we use a graph mapping that was designed in our prior work~\cite{aco_mcsp} to solve the MCSP problem using the Ant Colony Optimization technique~\cite{jour_dorigo}. Here we exploit this graph to present an Integer Programming (IP) formulation for the problem. Then we implement the MIP formulation, conduct extensive experiments and compare the results with the state-of-the-art algorithms in the literature. As will be reported in a later section, the results clearly indicate that the IP formulation is accurate, effective and provides excellent results.

The rest of the paper is organized as follows. In Section~\ref{sec:literature} we present a brief literature review. Section~\ref{sec:pre} presents the notations and definitions needed to understand the concept presented in the paper later. In Section~\ref{mip} we present the Integer Programming formulation for the MCSP problem. We present our experimental results in Section~\ref{exp}. Finally, we briefly conclude in Section~\ref{con}.
\section{Related Works}\label{sec:literature}

The 1-MCSP is essentially the breakpoint distance problem \cite{chromosome} between two permutations which is to count the number of ordered pairs of symbols that are adjacent in the first string but not in the other; this problem is obviously solvable in polynomial time \cite{goldstein}. The 2-MCSP is known to be NP-hard and moreover  APX-hard in \cite{goldstein}. The authors in \cite{goldstein} also presented several approximation algorithms. Chen et al. \cite{chen} studied the problem called the Signed Reversal Distance with Duplicates (SRDD), which is a generalization of MCSP. Furthermore, they gave a 1.5-approximation algorithm for 2-MCSP. In \cite{peter}, the author analyzed the fixed-parameter tractability of MCSP considering different parametrs.
In \cite{jiang}, the authors investigated $k$-MCSP along with two other variants, namely, $MCSP^c$, where the alphabet size is at most $c$ and $x$-balanced MCSP, which requires that the length of the blocks must be within the range $(n/d - x, n/d + x)$,
where $d$ is the number of blocks in the optimal common partition and $x$ is a constant integer. They showed that
$MCSP^c$ is NP-hard when $c \geq 2$. As for $k$-MCSP, they presented an FPT
algorithm which runs in $O^*((d!)^{2k})$ time.

Chrobak et al. \cite{chrobak} analyzed a natural greedy heuristic for MCSP: iteratively, at each step, it extracts a longest common substring from the input strings. They showed that for 2-MCSP, the approximation ratio (for the greedy heuristic) is exactly 3.
They also proved that for 4-MCSP the ratio would be $\log n$ and for the general MCSP, between $\Omega(n^{0.43})$ and $O(n^{0.67})$.

In our prior work \cite{aco_mcsp} we have developed a meta-heuristc algorithm, namely, MAX-MIN ant system to solve the MCSP problem. In particular, in \cite{aco_mcsp} we have mapped MCSP into a graph, namely, the common substring graph. MAX-MIN Ant System has been implemented over this graph. Very recently in \cite{hm_blum}, the authors have proposed an iterative probabilistic tree search algorithm for solving this problem. The algorithm is an iterative probabilistic variant of the greedy algorithm \cite{chrobak}. The authors have tested their approach with the dataset introduced in \cite{aco_mcsp}.

\section{Preliminaries}\label{sec:pre}

In this section, we present some definitions and notations that are used throughout the paper. Two strings $ (X,Y) $, each of length $n$, over an alphabet $\sum $ are called \emph{related} if every letter appears the same number of times in each of them. A block $B=([id,i,j])$, $0\leq i\leq j < n$, of a string $S$ is a data structure having three fields: $id$ is an identifier of $S$ and the starting and ending positions of the block in $S$ are represented by $i$ and $j$, respectively. Naturally, the \emph{length} of a block $[id,i,j]$ is $(j-i+1)$. We use $substring([id,i,j])$ to denote a substring of $S$ induced by the block $[id,i,j]$. 

For example, if we have two strings $(X,Y)$ = (``abcdab'',``bcdaba''), then $[0,0,1]$ and $[0,4,5]$ both represent the substring ``ab'' of $X$. In other words, $substring([0,0,1]) = substring([0,4,5]) =$ ``ab''. We say a block $B$ matches with another block $B'$ with the same or different id if the two blocks represent the same substrings. Given a block, $B$ along with a list of blocks $l_b$, we define $matchList(l_b,B)$ as a list of blocks of $l_b$ those match with $B$. For the above example, Let a list of blocks be $l_b$ = $\{[0,0,1],[0,1,1],[0,4,5]\}$ and $B = [0,0,1]$, then the $matchList(l_b,B)$ = $\{[0,0,1],[0,4,5]\}$.

We use the notion of a common substring graph as introduced in \cite{aco_mcsp}. The following definitions are borrowed from~\cite{aco_mcsp}. A common substring graph, $G_{cs}(V,E,id(X))$ of two strings $(X,Y)$ as follows. Here $V$ is the vertex set of the graph and $E$ is the edge set. Vertices are the positions of string $X$, i.e., for each $v \in V$, $v \in [0,|X|-1]$. Two vertices $v_i \leq v_j$ are connected with and edge, i.e, $(v_i, v_j) \in E$, if the substring induced by the block $[id(X),v_i,v_j]$ matches some substring of $Y$. More formally, if $S_y$ denotes the set of all substrings of $Y$, we have:

$$
    (v_i,v_j) \in E \Leftrightarrow substring([id(X),v_i,v_j]) == s\ is\ not\ empty\ \   \exists{s\in Y_s}
$$

In other words, each edge in the edge set corresponds to a \emph{block} satisfying the above condition. For convenience, we will denote the edges as \emph{edge blocks} and use the list of edge blocks (instead of edges) to define the edgeset $E$. 

For example, suppose $(X,Y)$ = (``abcdba'',``abcdab''). Now consider the corresponding common substring graph. Then, we have vertex set, $V=\{0,1,2,3,4,5\}$ and edge set, $E$ = $E = \{[0,0,0],[0,0,1],[0,1,1],[0,0,2],[0,0,3], [0,1,2], [0,1,3], [0,2,2],[0,2,3],$\\ $[0,3,3][0,4,4],[0,5,5]\}$. 

\section{IP Formulation}\label{mip}
Given two related strings $X$ and $Y$ each of length $n$, we create two graphs namely, $G_{cs}(V_1,E_1,id(X))$ and $G_{cs}(V_2,E_2,id(Y))$ of $(X,Y)$, where $V_1$ and $V_2$ are the vertex sets and $E_1$ and $E_2$
denote two edge block sets from the two graphs respectively. 
We define two sets of binary variables, namely, $x_{t1}$ and $y_{t2}$ where $t1 \in E_1$ and $t2 \in E_2$. We also
write $\delta_k(v)^-$ and $\delta_k(v)^+$ for the sets of incoming and outgoing edge blocks from
$E_k$ where $v \in V_k$ and $k \in \{1,2\}$. With the above settings, we develop the IP formulation for the MCSP as follows:

\begin{equation}
\label{obj}
    min \frac{(\sum_{t1 \in E_1} x_t + \sum_{t2 \in E_2} y_{t2})}{2}
\end{equation}

such that,

\begin{equation}
\label{cond1}
    \sum_{t1 \in E_1}{x_{t1}} = \sum_{t2 \in E_2}{y_{t2}}
\end{equation}

\begin{equation}
\label{cond2}
    \sum_{t1 \in \delta_1(0)^+} {x_{t1}}= 1
\end{equation}

\begin{equation}
\label{cond3}
    \sum_{t1 \in \delta_1(v)^-} {x_{t1}}= \sum_{t1 \in \delta_1(v+1)^+}{x_{t1}}\ \ \forall{v \in [0,n-1]}
\end{equation}

\begin{equation}
\label{cond4}
    \sum_{t2 \in \delta_2(0)^+} {y_{t2}}  = 1
\end{equation}

\begin{equation}
\label{cond5}
    \sum_{t2 \in \delta_2(v)^-} {y_{t2}}= \sum_{t \in \delta_2(v+1)^+}{y_{t2}} \ \ \forall{v \in [0,n-1]}
\end{equation}

\begin{equation}
\label{cond6}
    x_{t1} \leq \sum_{b \in matchList(E_2,t1)} y_b \ \ \forall{t1 \in E_1}
\end{equation}

\begin{equation}
\label{cond7}
    y_{t2} \leq \sum_{b \in matchList(E_1,t2)} x_b \ \ \forall{t2 \in E_2}
\end{equation}

\begin{equation}
\label{cond8}
    \sum_{b1 \in matchList(E_1,t1)} x_{b1} = \sum_{b2 \in matchList(E_2,t1)} y_{b2}  \ \ \forall{t1 \in E_1}
\end{equation}
$$
    x_{t1} \in \{0,1\}, y_{t2} \in \{0,1\}
$$
\subsection{Explanation of the Formulation}

\textbf{Objective function:} Eq. (\ref{obj}) is the objective function that is to be minimized. The function simply calculates the size of the partition.

\textbf{Equality constraint}: Eq. (\ref{cond1}) states that two partitions on the two substring graphs must be equal in size. In other words, the number of blocks in the factorization of the first string $X$ is equal to the number of blocks in the factorization of the second string $Y$.

\textbf{Factorization constraint}: Eqs. (\ref{cond2})-(\ref{cond3}) together imply that a unit flow enters at the source (the vertex labeled
with 0) and arrives at the sink (the vertex labeled with $n-1$) for string $X$. So, the
string is factorized. For string $Y$ the factorization is achieved in a similar fashion
by Eqs.(\ref{cond4})-(\ref{cond5}). These constraints ensure that the srtings get factorized by non-overlapping blocks.

\textbf{One to one match constraint}: 
We have two sets of blocks after the factorization. We must ensure that there is a one to one matching between the two sets of blocks. By matching we mean that, for each selected blocks (those with $x_t = 1$ where $t \in E_1$) of the first edge block set $E_1$, there must be one and only one corresponding selected block (with $y_t = 1$ where $t \in E_2$) with the same substring in the second edge block set $E_2$ and vice versa.  Eqs. (\ref{cond6})-(\ref{cond7}) achieve the matching between the two sets of blocks. Further, Eq. \ref{cond8} is needed for the one to one matching.

\section{Experiments}\label{exp}
We have conducted our experiments in a computer with Intel Core 2 Quad CPU 2.33 GHz. The available RAM was 4.00 GB. The operating system was Windows 7. The programming environment was Matlab. We have used SCIP (version 3.1.0) stand alone solver \cite{Achterberg2009} to solve the IP formulation (referred to as the IP algorithm henceforth).

\subsection{Datasets}
We have used the dataset used in our previous work \cite{aco_mcsp}. Notably, the same dataset has also been adopted to conduct experimental analysis and comparison by other researchers~\cite{hm_blum}. The data sets are briefly described below. There are two types of data: randomly generated DNA sequences and real gene sequences.
\subsubsection{Random DNA sequences:}
In  \cite{aco_mcsp}, we have generated $30$ random DNA sequences each of length at most 600 using \cite{shuffle}. The fraction of bases $A$, $T$, $G$ and $C$ is assumed to be 0.25 each. For each DNA sequence we shuffle it to create a new DNA sequence. The shuffling is done using the online toolbox \cite{shuffle}. The original random DNA sequence and its shuffled pair constitute a single input ($X,Y$) in our experiment. This dataset is divided into 3 groups. The first 10 (Group1) have lengths less than or equal to 200 bps (base-pairs), the next 10 (Group2) have lengths within $[201,400]$ and the rest 10 (Group3) have lengths within $[401,600]$ bps.

\subsubsection{Real Gene Sequences:}
We have collected the real gene sequence data used in \cite{aco_mcsp} and collected from the NCBI GenBank\footnote{http://www.ncbi.nlm.nih.gov}. This data correspond to the first 15 gene sequences of Bacterial Sequencing (part 14) whose lengths are within $[200,600]$. We will denote it by ``Real''.

\subsection{Implementation}

SCIP \cite{Achterberg2009} (version 3.1.0) is used to solve the IP formulation. We have used the stand alone solver. From the interface of the stand alone solver we have recorded the values of primal solution and the relative gap (gap = $|primal - dual|/MIN(|dual|,|primal|)$) periodically.  We have enforced a time limit of 15 minutes, 30 minutes and 60 minutes for the Group1, Group2 and Group3 dataset respectively. For the Real dataset, we have given a time according to the length of the sequence. If the length is not more than 200 bps, we have assigned 15 minutes. On the other hand if the sequence length is in between 200 and 400 bps, we have given 30 minutes. The rest of the instances are assigned 60 minutes time. All other parameters are left default.

\subsection{Results and Analysis}\label{results}
In an updated and extended version \cite{mcsp_aco_journal}\footnote{The preprint is available at~\cite{mcsp_aco_journal_arxiv}.} of our earlier work \cite{aco_mcsp}, MAX-MIN ACO (referred to as MMAS henceforth) has been compared with the greedy algorithm (referred to as Greedy henceforth) of \cite{chrobak}. In \cite{hm_blum}, the authors have compared their two versions of iterative probabilistic tree search (referred to TS1 and TS2 henceforth) with Greedy and MMAS. Here, we compare the IP algorithm with the above four algorithms, namely, MMAS \cite{mcsp_aco_journal_arxiv,mcsp_aco_journal,aco_mcsp}, Greedy \cite{chrobak} and both of TS1 and TS2 \cite{hm_blum}.

Tables (\ref{table:res_rand1})-(\ref{table:res_real}) present the results
for the Group1, Group2, Group3 and Real dataset respectively. We have taken the results of MMAS and Greedy from \cite{mcsp_aco_journal}. The results of TS1 and TS2 are taken from \cite{hm_blum}. As has been reported in \cite{mcsp_aco_journal}, in MMAS, for a particular DNA sequence, the experiment was run for 15 times and the average result is reported. The first column under any group reports the partition size computed by the greedy approach, the second column is the average partition size found by MMAS and the third column represents the average time in seconds for the MMAS solution. The fourth and fifth column report the partition size of TS1 and TS2 respectively. The sixth and seventh column represent the average time of TS1 and TS2 respectively. The eighth and ninth column report the partition size by IP algorithm and the time to achieve this result. We report the feasible primal solution value here. The relative gap of dual and primal solution is reported in the tenth column.
\begin{sidewaystable}
\begin{center}
\caption {Comparison among Greedy \cite{chrobak}, MMAS \cite{aco_mcsp}, TS1, TS2 \cite{hm_blum} and IP on random DNA sequences (Group 1, 200 bps). }
\label{table:res_rand1}
\scalebox{0.8}{
\begin{tabular}{ |l|l|l|l|l|l|l|l|l|l| }
  \hline
  \textbf{Greedy} & \textbf{MMAS(Avg.)} & \textbf{Avg. Time(MMAS)} & \textbf{TS1(Avg.)} & \textbf{TS2(Avg.)} & \textbf{Avg. Time(TS1)} & \textbf{Avg. Time(TS2)} & \textbf{IP} &  \textbf{Time(IP)} & \textbf{GAP}\\
  \hline
  46 & 42.87 & 114.62 & 42.50 & 42.30 & 197.36 & 285.63 & 41 & 22.80 & 0 \\
56 & 51.87 & 100.82 & 48.90 & 48.90 & 175.95 & 91.96 & 47 & 30.50 & 0 \\
62 & 57.00 & 207.53 & 56.00 & 56.00 & 253.47 & 123.97 & 54 & 200.00 & 0 \\
46 & 43.33 & 168.31 & 43.00 & 43.00 & 52.34 & 88.72 & 41 & 18.00 & 0 \\
44 & 42.93 & 42.71 & 41.00 & 41.00 & 124.53 & 30.48 & 40 & 16.00 & 0 \\
48 & 42.80 & 75.20 & 41.10 & 41.70 & 278.28 & 221.10 & 40 & 10.00 & 0 \\
65 & 60.60 & 131.95 & 60.80 & 61.00 & 106.12 & 21.34 & 55 & 65.00 & 0 \\
51 & 46.93 & 201.23 & 45.30 & 45.30 & 389.73 & 369.13 & 43 & 19.00 & 0 \\
46 & 45.53 & 172.68 & 43.00 & 43.00 & 247.70 & 166.10 & 42 & 16.00 & 0 \\
63 & 59.73 & 288.42 & 58.80 & 59.00 & 218.88 & 76.57 & 54 & 160.00 & 0 \\
  \hline

\end{tabular}
}
\end{center}
\end{sidewaystable}

\begin{sidewaystable}
\begin{center}
\caption {Comparison among Greedy \cite{chrobak}, MMAS \cite{aco_mcsp}, TS1, TS2 \cite{hm_blum} and IP on random DNA sequences (Group 2, 400 bps).}
\label{table:res_rand2}
\scalebox{0.8}{
\begin{tabular}{ |l|l|l|l|l|l|l|l|l|l| }
  \hline
  \textbf{Greedy} & \textbf{MMAS(Avg.)} & \textbf{Avg. Time(MMAS)} & \textbf{TS1(Avg.)} & \textbf{TS2(Avg.)} & \textbf{Avg. Time(TS1)} & \textbf{Avg. Time(TS2)} & \textbf{IP} &  \textbf{Time(IP)} & \textbf{GAP}\\
  \hline
  119 & 113.93 & 1534.10 & 112.80 & 112.10 & 236.02 & 270.48 & 99 & 432 & 3.95 \\
122 & 118.93 & 1683.11 & 115.60 & 115.60 & 471.67 & 466.58 & 104 & 1384 & 5.50 \\
114 & 112.53 & 1398.53 & 108.30 & 107.60 & 207.00 & 501.57 & 97 & 1600 & 2.91 \\
116 & 116.40 & 1739.35 & 112.40 & 112.40 & 291.36 & 206.16 & 101 & 500 & 3.68 \\
135 & 132.20 & 1814.73 & 128.70 & 129.50 & 373.68 & 379.25 & 115 & 1300 & 6.09 \\
108 & 106.07 & 1480.24 & 103.60 & 103.20 & 353.94 & 229.14 & 95 & 500 & 7.77 \\
108 & 98.40 & 1295.25 & 96.90 & 96.70 & 327.40 & 318.24 & 88 & 850 & 6.33 \\
123 & 118.40 & 1125.24 & 115.10 & 115.30 & 369.12 & 305.54 & 104 & 430 & 5.32 \\
124 & 119.47 & 1044.41 & 114.80 & 114.50 & 235.29 & 281.77 & 104 & 100 & 5.39 \\
105 & 101.87 & 1360.15 & 98.60 & 98.70 & 162.48 & 308.61 & 89 & 560 & 3.49 \\
\hline
\end{tabular}
}
\end{center}
\end{sidewaystable}

\begin{sidewaystable}
\begin{center}
\caption {Comparison among Greedy \cite{chrobak}, MMAS \cite{aco_mcsp}, TS1, TS2 \cite{hm_blum} and IP on random DNA sequences (Group 3, 600 bps).}
\label{table:res_rand3}
\scalebox{0.8}{
\begin{tabular}{ |l|l|l|l|l|l|l|l|l|l| }
  \hline
  \textbf{Greedy} & \textbf{MMAS(Avg.)} & \textbf{Avg. Time(MMAS)} & \textbf{TS1(Avg.)} & \textbf{TS2(Avg.)} & \textbf{Avg. Time(TreeSearch1)} & \textbf{Avg. Time(TreeSearch2)} & \textbf{IP} &  \textbf{Time(IP)} & \textbf{GAP}\\
  \hline
 182 & 179.93 & 1773.04 & 172.90 & 172.9 & 196.92 & 434.40 & 155 & 415 & 8.10 \\
175 & 176.20 & 3966.83 & 170.80 & 170.7 & 390.75 & 396.32 & 152 & 2105 & 5.59 \\
196 & 187.87 & 1589.30 & 186.30 & 186.8 & 361.04 & 446.02 & 161 & 644 & 5.75 \\
192 & 184.27 & 2431.16 & 181.00 & 180.5 & 335.72 & 423.51 & 159 & 1755 & 7.38 \\
176 & 171.53 & 1224.89 & 165.00 & 164.7 & 399.88 & 437.65 & 153 & 2400 & 9.17 \\
170 & 163.47 & 1826.14 & 164.40 & 164.4 & 427.94 & 506.95 & 145 & 1100 & 8.15 \\
173 & 168.47 & 1802.17 & 162.40 & 162.6 & 488.96 & 474.20 & 146 & 1400 & 8.46 \\
185 & 176.33 & 1838.56 & 172.40 & 171.9 & 316.08 & 376.09 & 151 & 515 & 6.86 \\
174 & 172.80 & 4897.47 & 170.60 & 170.4 & 365.62 & 368.80 & 154 & 2700 & 9.25 \\
171 & 167.20 & 1886.21 & 162.50 & 162.3 & 346.42 & 483.19 & 146 & 2600 & 8.61 \\
  \hline

\end{tabular}
}
\end{center}
\end{sidewaystable}

\begin{sidewaystable}
\begin{center}
\caption {Comparison among Greedy \cite{chrobak}, MMAS \cite{aco_mcsp}, TS1, TS2 \cite{hm_blum} and IP on Real }
\label{table:res_real}
\scalebox{0.7}{
\begin{tabular}{ |l|l|l|l|l|l|l|l|l|l| }
  \hline
  \textbf{Greedy} & \textbf{MMAS(Avg.)} & \textbf{Avg. Time(MMAS)} & \textbf{TS1(Avg.)} & \textbf{TS2(Avg.)} & \textbf{Avg. Time(TS1)} & \textbf{Avg. Time(TS2)} & \textbf{IP} &  \textbf{Time(IP)} & \textbf{GAP}\\
  \hline
95 & 87.67 & 863.81 & 87.80 & 87.30 & 314.42 & 314.42 & 78 & 500 & 2.90 \\
161 & 156.33 & 1748.34 & 154.50 & 155.50 & 384.93 & 424.61 & 136 & 2280 & 7.18 \\
121 & 117.07 & 1823.49 & 113.80 & 113.80 & 268.66 & 430.52 & 104 & 1620 & 6.01 \\
173 & 164.87 & 1823.01 & 160.60 & 160.30 & 360.61 & 436.76 & 142 & 2400 & 5.18 \\
172 & 171.07 & 2210.15 & 167.80 & 167.60 & 521.06 & 375.17 & 149 & 1195 & 7.78 \\
153 & 146.00 & 1953.84 & 144.90 & 144.10 & 212.69 & 365.66 & 127 & 1900 & 6.06 \\
140 & 141.00 & 2439.03 & 133.00 & 132.50 & 425.30 & 286.02 & 120 & 850 & 6.37 \\
134 & 133.13 & 1406.80 & 128.70 & 128.90 & 414.49 & 482.46 & 117 & 1740 & 8.12 \\
149 & 147.53 & 2547.52 & 142.60 & 142.70 & 314.78 & 330.21 & 128 & 33 & 6.59 \\
151 & 150.53 & 1619.64 & 145.30 & 145.60 & 465.11 & 274.35 & 128 & 3171 & 4.64 \\
126 & 125.00 & 1873.39 & 121.60 & 121.70 & 464.24 & 331.92 & 112 & 1524 & 7.00 \\
143 & 139.13 & 2473.25 & 139.00 & 139.40 & 360.15 & 256.56 & 123 & 785 & 7.90 \\
180 & 181.53 & 2931.67 & 173.20 & 173.20 & 417.10 & 455.36 & 155 & 1427 & 6.63 \\
152 & 149.33 & 2224.40 & 147.80 & 147.30 & 367.72 & 465.47 & 134 & 670 & 7.85 \\
157 & 161.60 & 1739.61 & 153.20 & 153.10 & 313.26 & 389.08 & 144 & 3400 & 7.21 \\
  \hline

\end{tabular}
}
\end{center}
\end{sidewaystable}

Tables (\ref{table:res_rand1_stat})-(\ref{table:res_real_stat}) report the improvement achieved by IP algorithm over Greedy \cite{chrobak}, MMAS \cite{aco_mcsp}, TS1, TS2 \cite{hm_blum} for the Group1, Group2, Group3 and Real dataset, respectively. Some statistical test results are also reported in these tables. The first four columns of each table here represent the differences in the partition size between IP algorithm and the other four approaches (Greedy, MMAS, TS1 and TS2 respectively). A negative (positive) result indicates that the IP algorithm is better (worse) than the other algorithm by that amount. The fifth to seventh columns of each table report the result of student's \emph{t}-test between the MMAS and IP algorithm. For each instance we have a vector of 15 sample observations from MMAS. The result of IP algorithm is replicated 15 times. The \emph{t}-test is performed on these two vectors. The fifth column reports the \emph{t}-stat value. The sixth column represent the \emph{p}-value of the test. A positive \emph{t}-stat value with a low \emph{p}-value indicates improvement. The seventh column reports a test decision for the null hypothesis that the data in two vectors (MMAS and IP results) come from independent random samples from normal distributions with equal means and equal but unknown variances, using the two-sample \emph{t}-test. We use +,-,$\approx$ to indicate a better (a positive \emph{t}-stat value), worse (a negative \emph{t}-stat value ) or equal (higher \emph{p}-value than the significance level) result than MMAS. Here the significance level is 5\%. 

Similarly, The next three columns report the \emph{t}-test results between TS1 and IP algorithm. Finally, the last three columns represent the statistics result between TS2 and IP algorithm. The two pair sampled \emph{t}-stat value for TS1 and TS2 is obtained by the average and standard deviation of the 10 independent runs of TS1 and TS2 reported in \cite{hm_blum}.

IP algorithm is not a stochastic algorithm. So it has always a standard deviation of zero. As a result the application of \emph{t}-test is dependent on the stochastic nature of the other algorithms. If the comparing algorithm also has a zero standard deviation than we will not get any \emph{t}-test result. Those cases are shown in the tables as ``NA''.  As it can be seen from the tables all of the differences are negative. That means our result is better than all other approaches. The result is also justified statistically. All of the \emph{p}-values are zero and all of the \emph{t}-stat values are positive.

\begin{table}
\begin{center}
\caption {Differences and Statistical Results (Group1). Here Diff1 = IP-Greedy, Diff2 = IP - MMAS, Diff3 = IP - TS1, Diff4 = IP-TS2}
\label{table:res_rand1_stat}
\scalebox{0.8}{
\begin{tabular}{ |l|l|l|l|l|l|l|l|l|l|l|l|l| }
  \hline
  \textbf{Diff1} & \textbf{Diff2} & \textbf{Diff3} & \textbf{Diff4} & \multicolumn{3}{|l|}{IP vs. MMAS} & \multicolumn{3}{|l|}{IP vs. TS1} & \multicolumn{3}{|l|}{IP vs. TS2} \\
  \hline
   & & & & \textbf{\emph{t}-stat} & \textbf{\emph{p}-value} & \textbf{significance} & \textbf{\emph{t}-stat} & \textbf{\emph{p}-value} & \textbf{significance} & \textbf{\emph{t}-stat} & \textbf{\emph{p}-value} & \textbf{significance}\\
  \hline
-5 & -3.13 & -1.50 & -1.30 & 20.55 & 0.00 & + & 8.95 & 0 & + & 8.56 & 0 & + \\
-9 & -4.13 & -1.90 & -1.90 & 36.50 & 0.00 & + & 18.78 & 0 & + & NA & NA & NA \\
-8 & -5.00 & -2.00 & -2.00 & 17.75 & 0.00 & + & NA & NA & NA & NA & NA & NA \\
-5 & -2.67 & -2.00 & -2.00 & 18.52 & 0.00 & + & NA & NA & NA & NA & NA & NA \\
-4 & -1.07 & -1.00 & -1.00 & 44.00 & 0.00 & + & NA & NA & NA & NA & NA & NA \\
-8 & -5.20 & -1.10 & -1.70 & 26.19 & 0.00 & + & 10.87 & 0 & + & 11.20 & 0 & + \\
-10 & -4.40 & -5.80 & -6.00 & 42.77 & 0.00 & + & 43.67 & 0 & + & NA & NA & NA \\
-8 & -4.07 & -2.30 & -2.30 & 33.28 & 0.00 & + & 15.15 & 0 & + & 15.15 & 0 & + \\
-4 & -0.47 & -1.00 & -1.00 & 26.50 & 0.00 & + & NA & NA & NA & NA & NA & NA \\
-9 & -3.27 & -4.80 & -5.00 & 31.55 & 0.00 & + & 36.14 & 0 & + & NA & NA & NA \\

  \hline

\end{tabular}
}
\end{center}
\end{table}

\begin{table}
\begin{center}
\caption {Differences and Statistical Results (Group2). Here Diff1 = IP-Greedy, Diff2 = IP - MMAS, Diff3 = IP - TS1, Diff4 = IP-TS2}
\label{table:res_rand2_stat}
\scalebox{0.8}{
\begin{tabular}{ |l|l|l|l|l|l|l|l|l|l|l|l|l| }
  \hline
  \textbf{Diff1} & \textbf{Diff2} & \textbf{Diff3} & \textbf{Diff4} & \multicolumn{3}{|l|}{IP vs. MMAS} & \multicolumn{3}{|l|}{IP vs. TS1} & \multicolumn{3}{|l|}{IP vs. TS2} \\
  \hline
   & & & & \textbf{\emph{t}-stat} & \textbf{\emph{p}-value} & \textbf{significance} & \textbf{\emph{t}-stat} & \textbf{\emph{p}-value} & \textbf{significance} & \textbf{\emph{t}-stat} & \textbf{\emph{p}-value} & \textbf{significance}\\
  \hline
-20 & -5.07 & -13.80 & -13.10 & 43.34 & 0.00 & + & 103.90 & 0.00 & + & 55.98 & 0.00 & + \\
-18 & -3.07 & -11.60 & -11.60 & 60.17 & 0.00 & + & 52.40 & 0.00 & + & 73.36 & 0.00 & + \\
-17 & -1.47 & -11.30 & -10.60 & 72.15 & 0.00 & + & 53.33 & 0.00 & + & 64.46 & 0.00 & + \\
-15 & 0.40 & -11.40 & -11.40 & 80.95 & 0.00 & + & 51.50 & 0.00 & + & 51.50 & 0.00 & + \\
-20 & -2.80 & -13.70 & -14.50 & 50.46 & 0.00 & + & 37.35 & 0.00 & + & 53.94 & 0.00 & + \\
-13 & -1.93 & -8.60 & -8.20 & 48.50 & 0.00 & + & 52.30 & 0.00 & + & 41.16 & 0.00 & + \\
-20 & -9.60 & -8.90 & -8.70 & 32.43 & 0.00 & + & 87.95 & 0.00 & + & 41.06 & 0.00 & + \\
-19 & -4.60 & -11.10 & -11.30 & 75.69 & 0.00 & + & 109.69 & 0.00 & + & 53.33 & 0.00 & + \\
-20 & -4.53 & -10.80 & -10.50 & 56.51 & 0.00 & + & 54.21 & 0.00 & + & 34.23 & 0.00 & + \\
-16 & -3.13 & -9.60 & -9.70 & 67.05 & 0.00 & + & 58.38 & 0.00 & + & 63.90 & 0.00 & + \\
  \hline

\end{tabular}
}
\end{center}
\end{table}

\begin{table}
\begin{center}
\caption {Differences and Statistical Results (Group3). Here Diff1 = IP-Greedy, Diff2 = IP - MMAS, Diff3 = IP - TS1, Diff4 = IP-TS2}
\label{table:res_rand3_stat}
\scalebox{0.8}{
\begin{tabular}{ |l|l|l|l|l|l|l|l|l|l|l|l|l| }
  \hline
  \textbf{Diff1} & \textbf{Diff2} & \textbf{Diff3} & \textbf{Diff4} & \multicolumn{3}{|l|}{IP vs. MMAS} & \multicolumn{3}{|l|}{IP vs. TS1} & \multicolumn{3}{|l|}{IP vs. TS2} \\
  \hline
   & & & & \textbf{\emph{t}-stat} & \textbf{\emph{p}-value} & \textbf{significance} & \textbf{\emph{t}-stat} & \textbf{\emph{p}-value} & \textbf{significance} & \textbf{\emph{t}-stat} & \textbf{\emph{p}-value} & \textbf{significance}\\
  \hline
-27 & -2.07 & -17.90 & -17.90 & 56.47 & 0.00 & + & 47.17 & 0.00 & + & 64.32 & 0.00 & + \\
-23 & 1.20 & -18.80 & -18.70 & 108.74 & 0.00 & + & 48.33 & 0.00 & + & 123.20 & 0.00 & + \\
-35 & -8.13 & -25.30 & -25.80 & 140.00 & 0.00 & + & 119.41 & 0.00 & + & 129.50 & 0.00 & + \\
-33 & -7.73 & -22.00 & -21.50 & 213.78 & 0.00 & + & 74.01 & 0.00 & + & 79.99 & 0.00 & + \\
-23 & -4.47 & -12.00 & -11.70 & 78.41 & 0.00 & + & 80.74 & 0.00 & + & 45.12 & 0.00 & + \\
-25 & -6.53 & -19.40 & -19.40 & 38.73 & 0.00 & + & 87.64 & 0.00 & + & 63.25 & 0.00 & + \\
-27 & -4.53 & -16.40 & -16.60 & 73.29 & 0.00 & + & 61.74 & 0.00 & + & 62.49 & 0.00 & + \\
-34 & -8.67 & -21.40 & -20.90 & 120.17 & 0.00 & + & 130.14 & 0.00 & + & 48.24 & 0.00 & + \\
-20 & -1.20 & -16.60 & -16.40 & 46.45 & 0.00 & + & 100.95 & 0.00 & + & 61.74 & 0.00 & + \\
-25 & -3.80 & -16.50 & -16.30 & 146.46 & 0.00 & + & 61.39 & 0.00 & + & 62.86 & 0.00 & + \\
  \hline

\end{tabular}
}
\end{center}
\end{table}


\begin{table}
\begin{center}
\caption {Difference and Statistical Result (Real). Here Diff1 = IP-Greedy, Diff2 = IP - MMAS, Diff3 = IP - TS1, Diff4 = IP-TS2}
\label{table:res_real_stat}
\scalebox{0.8}{
\begin{tabular}{ |l|l|l|l|l|l|l|l|l|l|l|l|l|}
  \hline
  \textbf{Diff1} & \textbf{Diff2} & \textbf{Diff3} & \textbf{Diff4} & \multicolumn{3}{|l|}{IP vs. MMAS} & \multicolumn{3}{|l|}{IP vs. TS1} & \multicolumn{3}{|l|}{IP vs. TS2} \\
  \hline
  & & & & \textbf{\emph{t}-stat} & \textbf{\emph{p}-value} & \textbf{significance} & \textbf{\emph{t}-stat} & \textbf{\emph{p}-value} & \textbf{significance} & \textbf{\emph{t}-stat} & \textbf{\emph{p}-value} & \textbf{significance}\\
  \hline
-17 & -7.33 & -9.80 & -9.30 & 76.73 & 0.00 & + & 73.79 & 0.00 & + & 43.89 & 0.00 & + \\
-25 & -4.67 & -18.50 & -19.50 & 33.51 & 0.00 & + & 110.38 & 0.00 & + & 116.35 & 0.00 & + \\
-17 & -3.93 & -9.80 & -9.80 & 57.27 & 0.00 & + & 49.19 & 0.00 & + & 49.19 & 0.00 & + \\
-31 & -8.13 & -18.60 & -18.30 & 74.60 & 0.00 & + & 70.02 & 0.00 & + & 49.89 & 0.00 & + \\
-23 & -0.93 & -18.80 & -18.60 & 77.71 & 0.00 & + & 57.72 & 0.00 & + & 54.97 & 0.00 & + \\
-26 & -7.00 & -17.90 & -17.10 & 56.20 & 0.00 & + & 70.76 & 0.00 & + & 73.07 & 0.00 & + \\
-20 & 1.00 & -13.00 & -12.50 & 107.59 & 0.00 & + & 43.73 & 0.00 & + & 55.67 & 0.00 & + \\
-17 & -0.87 & -11.70 & -11.90 & 34.57 & 0.00 & + & 77.08 & 0.00 & + & 66.02 & 0.00 & + \\
-21 & -1.47 & -14.60 & -14.70 & 50.25 & 0.00 & + & 88.79 & 0.00 & + & 69.38 & 0.00 & + \\
-23 & -0.47 & -17.30 & -17.60 & 54.63 & 0.00 & + & 66.72 & 0.00 & + & 107.03 & 0.00 & + \\
-14 & -1.00 & -9.60 & -9.70 & 50.35 & 0.00 & + & 58.38 & 0.00 & + & 63.90 & 0.00 & + \\
-20 & -3.87 & -16.00 & -16.40 & 50.15 & 0.00 & + & 61.70 & 0.00 & + & 74.09 & 0.00 & + \\
-25 & 1.53 & -18.20 & -18.20 & 75.80 & 0.00 & + & 50.49 & 0.00 & + & 62.56 & 0.00 & + \\
-18 & -2.67 & -13.80 & -13.30 & 46.00 & 0.00 & + & 69.27 & 0.00 & + & 62.77 & 0.00 & + \\
-13 & 4.60 & -9.20 & -9.10 & 54.88 & 0.00 & + & 31.62 & 0.00 & + & 50.49 & 0.00 & + \\
  \hline

\end{tabular}
}
\end{center}
\end{table}

\begin{figure}

  \includegraphics[width=1\textwidth]{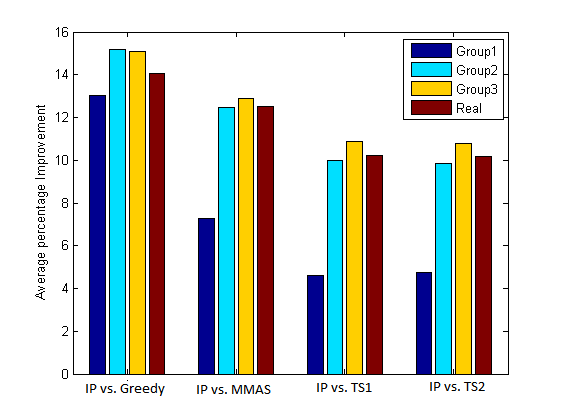}
  \caption{Average percentage of improvement of IP algorithm over Greedy, MMAS, TS1 and TS2 approaches}
  \label{avg_improvement}
\end{figure}

Figure \ref{avg_improvement} shows the average percentage of improvement of IP Algorithm over the other four approaches. The percentage of improvement is between 4\% to 16\%.

\subsection{Running Time}
\begin{figure}

  \begin{center}
  \includegraphics[width=1.3\textwidth]{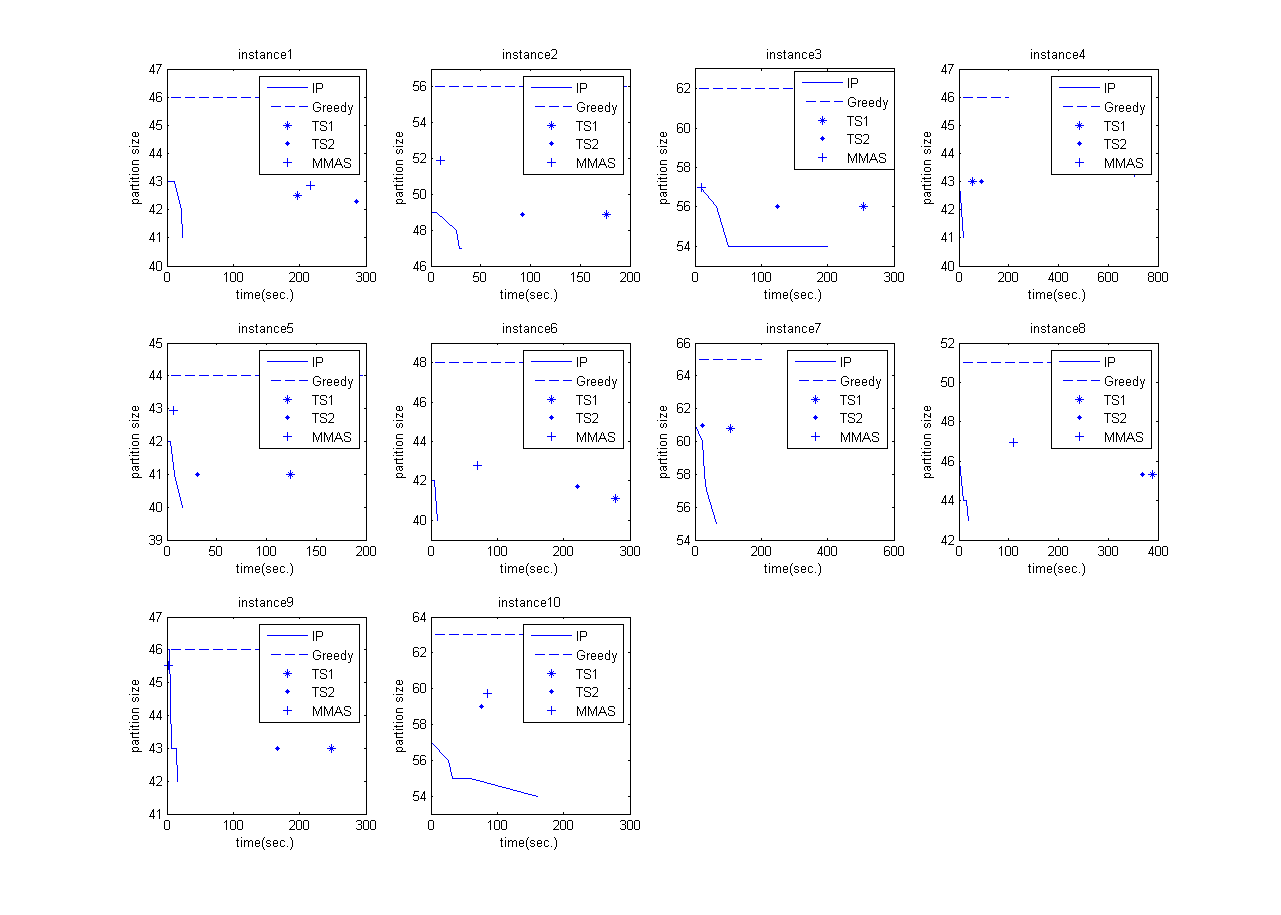}
  \caption{Running time comparison (Group1)}
  \label{runtime_200}
  \end{center}
\end{figure}

\begin{figure}

  \includegraphics[width=1.3\textwidth]{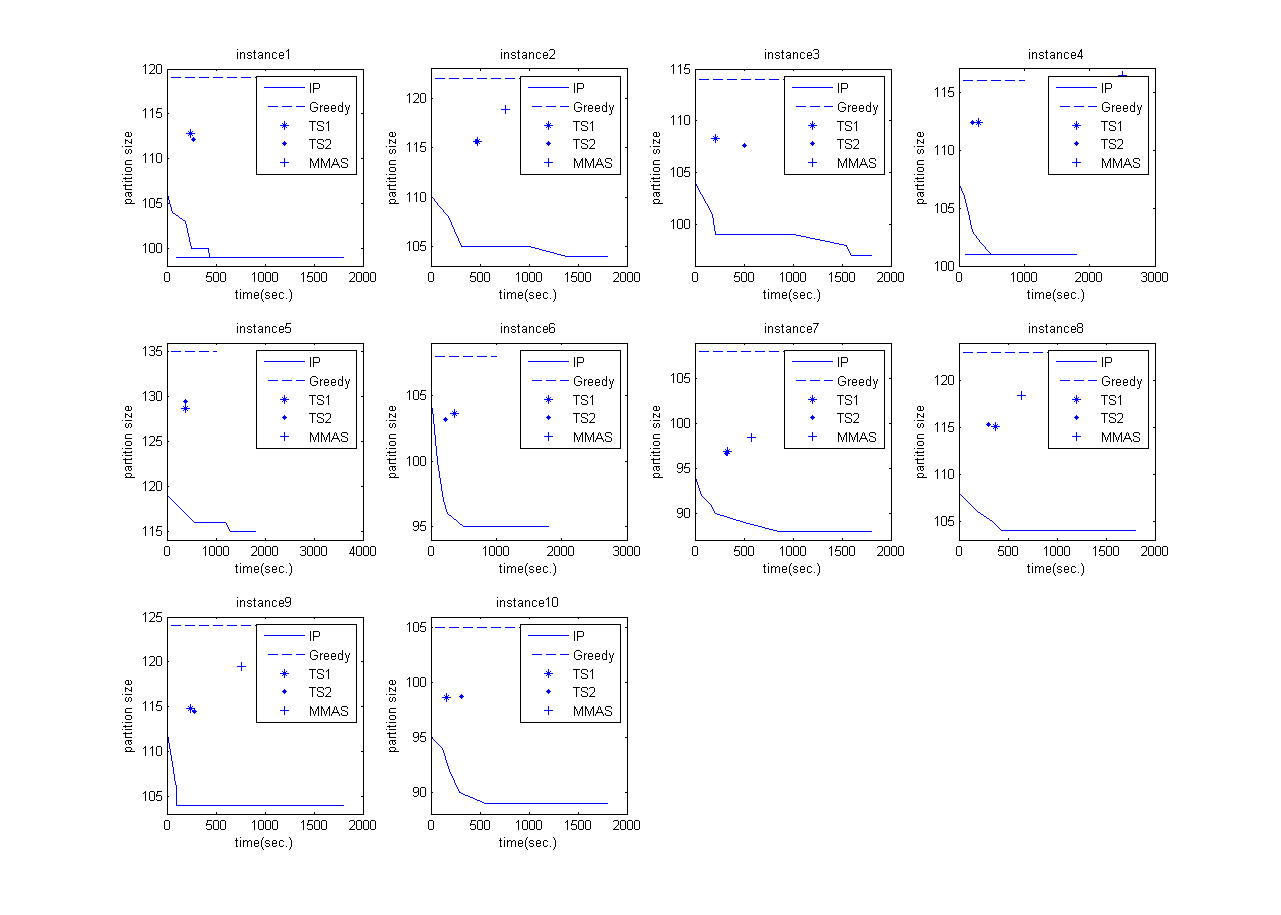}
  \caption{Running time comparison (Group2)}
  \label{runtime_400}
\end{figure}

\begin{figure}

  \includegraphics[width=1.3\textwidth]{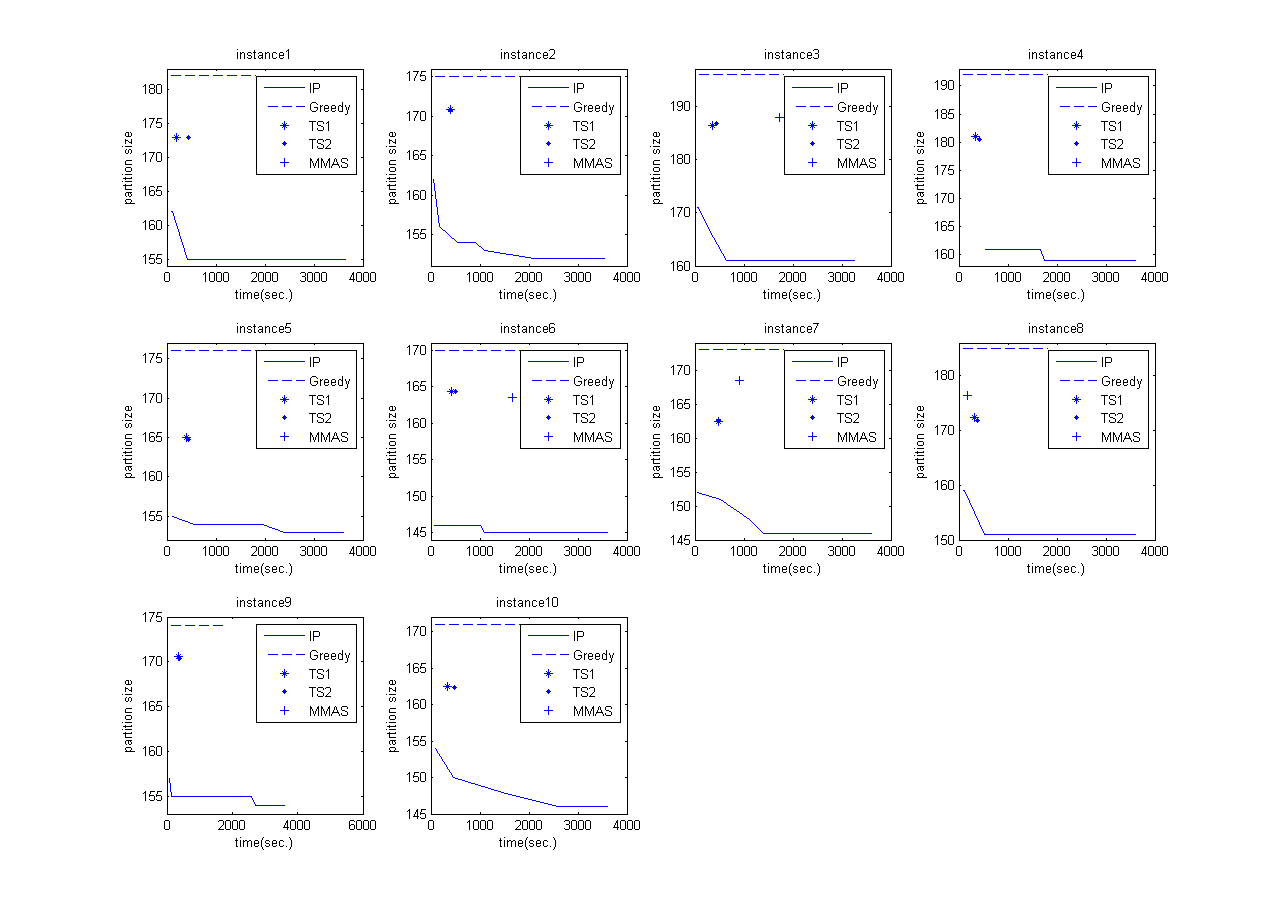}
  \caption{Running time comparison (Group3)}
  \label{runtime_600}
\end{figure}

\begin{figure}

  \includegraphics[width=1.2\textwidth]{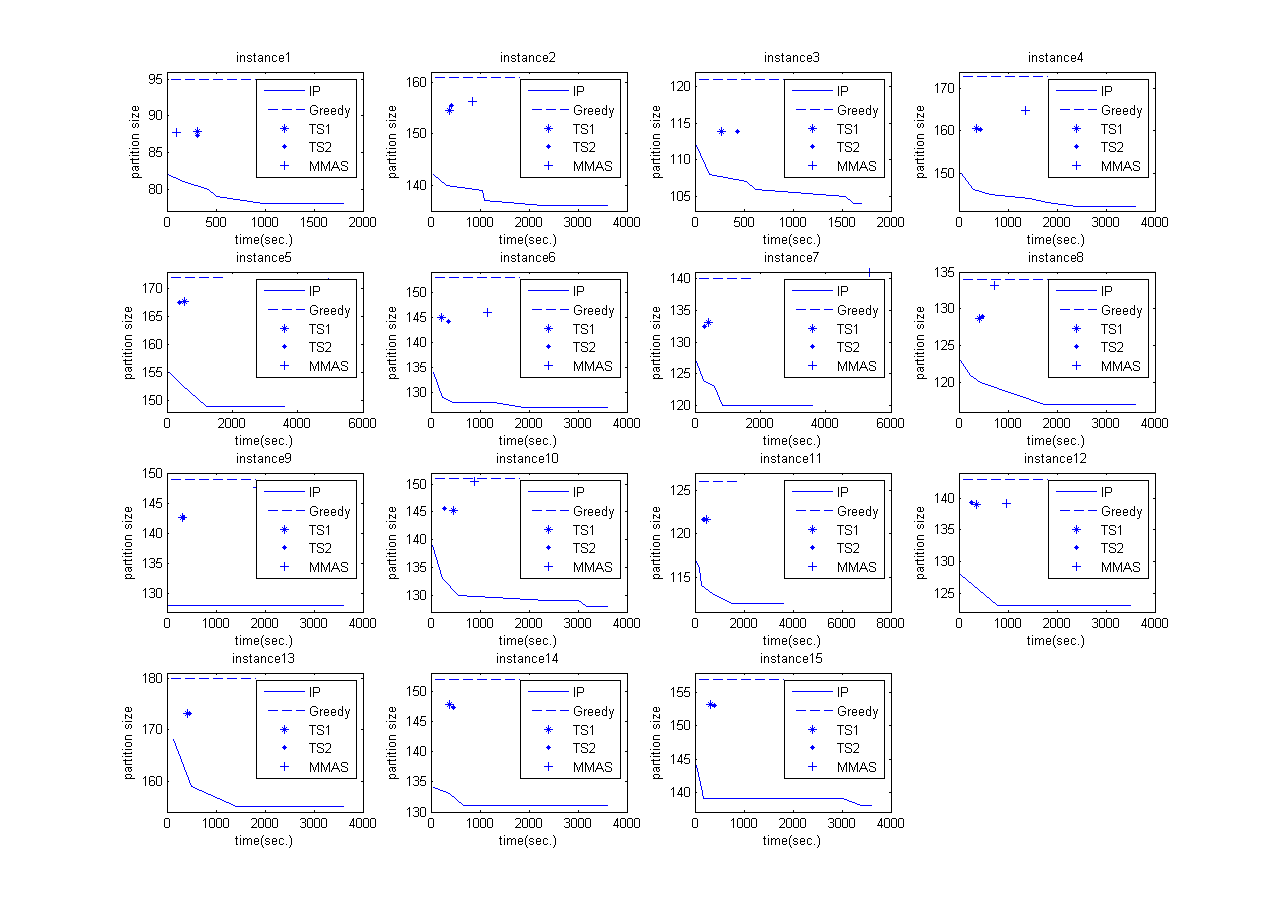}
  \caption{Running time comparison (Real)}
  \label{runtime_real}
\end{figure}

In the previous section, we have shown that the IP algorithm provides much better partition size. In this section we will explore the runtime of the IP algorithm and compare it with the other four approaches. The runtime of TS1 and TS2 is taken from \cite{Achterberg2009}. The runtime of MMAS is taken from \cite{mcsp_aco_journal}. The Greedy algorithm is very fast. It gives the output within 2 minutes. So, in the analysis, we will assume that the output of Greedy algorithm is readily available even at the beginning of simulation. We have recorded the primal solution (partition size) of the IP algorithm periodically. Figures (\ref{runtime_200})-(\ref{runtime_real}) show the detailed runtime comparison among the algorithms for Group1, Group2, Group3 and Real datasets respectively. For each group we have shown the partition size dynamics with respect to the time. The three points in each of the figure (``*'',``.'',``+'') are the plots of partition size vs. the average time needed to achieve that partition size for the TS1, TS2 and MMAS approach respectively (from Tables (\ref{table:res_rand1})-(\ref{table:res_rand3}),\ref{table:res_real}). The broken line represents the Greedy partition size.

Although the reported time (in tables (\ref{table:res_rand1})-(\ref{table:res_real})) of IP algorithm is higher than Greedy, TS1 and TS2 approaches in some instances but from the Figures (\ref{runtime_200})-(\ref{runtime_real}), it can be easily observed that the IP algorithm reaches to better solutions much earlier. From the figures it is clear that the IP algorithm is better than Greedy at any stage of time. Even if we stop the IP algorithm at or earlier than the average runtime of MMAS, TS1 or TS2, the IP algorithm provides better solutions.

\section{Conclusion}\label{con}
In this paper, we have presented an Integer Programming formulation for the MCSP problem. We have conducted extensive experiments and compare the results with the state of the art algorithms in the literature. The results clearly indicate that the IP formulation is accurate, effective and provides excellent results.
\bibliographystyle{abbrv}
\bibliography{Bibliography}

\end{document}